\documentstyle[bo99,epsfig]{article}
\newcommand {\BS}{{{\it Beppo}SAX}\ \ignorespaces}
\def\gta{ \lower .75ex \hbox{$\sim$} \llap{\raise .27ex \hbox{$>$}} }
\def\lta{ \lower .75ex\hbox{$\sim$} \llap{\raise .27ex \hbox{$<$}} }

\title{A self-consistent test of Comptonization models using a long
BeppoSAX observation of NGC 5548}

\author{P.O. Petrucci $^1$, F. Haardt$^2$, L. Maraschi$^1$,
P. Grandi$^3$, G. Matt$^6$, F. Nicastro$^{3,4,5}$, L. Piro$^3$,
G.C. Perola$^6$, A. De Rosa$^3$} 

\affil{1) Osservatorio Astronomico di Brera, Milano, Italy, 2)
Universit\'a dell'Insubria, Como, Italy, 3) IAS/CNR, Roma, Italy, 4) CfA,
Cambridge Ma., USA, 5) Osservatorio Astronomico di Roma, Roma, Italy, 6)
Universit\'a degli Studi ``Roma 3'', Roma, Italy} 

\begin{document}

\maketitle

\begin{abstract}
We test accurate models of Comptonization spectra over the high quality
data of the \BS long look at NGC 5548.
The data are well represented by a plane parallel corona with an
inclination angle of 30$^{\circ}$, a soft photon temperature of 5 eV and
a hot plasma temperature and optical depth of $kT_{\rm e}\simeq$ 360 keV
and $\tau\simeq$ 0.1, respectively. If energy balance applies, such
values
suggest that a more ``photon-starved'' geometry (e.g. a hemispheric
region) is necessary. The spectral softening detected
during a flare,  appears
to be associated to a decrease of the heating--to--cooling ratio,
indicating a geometric and/or energetic modification of the disk plus
corona system.
The hot plasma temperature derived with the models above is 
significantly higher than that obtained 
fitting the same data with a power law plus high energy cut off 
model for the continuum. This is due to
the fact that in anisotropic geometries Comptonization spectra show
"intrinsic" curvature which moves the fitted  high energy cut-off to higher
energies.

\end{abstract}

\section{Introduction}
The X-ray emission of Seyfert galaxies is commonly believed to be
produced by Compton scattering of soft photons on a (thermal or
non-thermal) population of hot electrons. The non--detection of Seyferts
by Comptel and the high energy cut-offs indicated by OSSE have focused
attention on thermal models (e.g. Sunyaev \& Titarchuk, 1980). In this
case the X--ray spectral shape is mainly determined by the temperature
$\Theta=kT_{\rm e}/m_{\rm e}c^2$ and the optical depth $\tau$ of the
scattering electrons, while the cut--off energy is related essentially to
$\Theta$. Moreover, if the Comptonizing region and the source of seed
photons are {\it coupled}, one can write an energy balance equation for
the hot coronal plasma which determines a roughly constant value of the
Compton parameter $y\simeq 4\Theta \tau \, (1+4\Theta)(1+\tau)$ (e.g.,
Svensson, 1996) and thus a one to one correpondence between $\theta$ and
$\tau$.  The required value of $y$ depends on the fraction $f$ of the
power dissipated in the corona and on geometry. In the following the
limiting case $f=1$ is considered.\\
\noindent
In the present work (Petrucci et al., 1999, hereafter P99), we test
Comptonization models over the high quality data of the (8 days) \BS long
look at the Seyfert I galaxy NGC 5548, deriving constraints on the
physical parameters and geometry of the source.
These data have already been studied in detail by Nicastro et al. (1999),
modelling the continuum with a cut-off power law.

\begin{figure}[h]
\centerline{\psfig{file=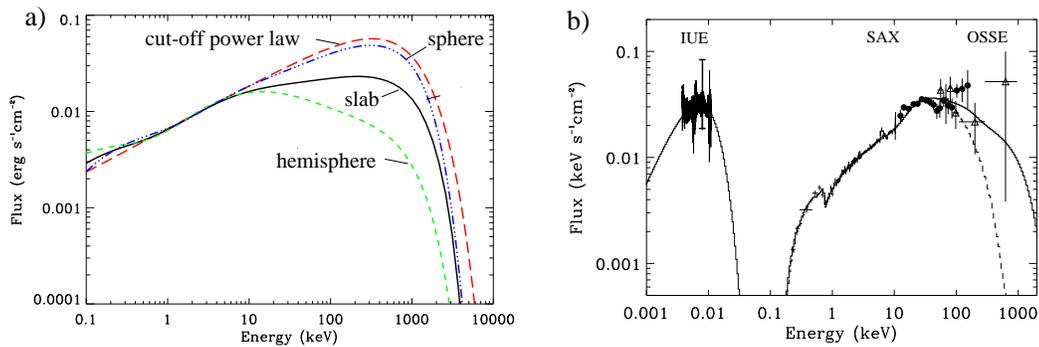, width=5cm,angle=-90}}
\caption{{\bf (a):} Comptonized models for different geometries assuming
$\Theta=0.7$. We have also over-plotted a cut--off power law with
$E_{\rm c}=2kT_{\rm e}$ in dashed line (cf. text for details).{\bf (b):}
\BS data set of NGC 5548 with non--simultaneous IUE and OSSE data (from
Magdziark et al., 1998) with the corresponding best fits Comptonization
model (in slab geometry, solid line) and simple cut--off power law model
(dashed line).}
\vspace*{-0.5cm}
\end{figure}

\section{The Comptonization Model}
\subsection{The anisotropy break}
In Fig 1a we show comptonized spectra computed for different geometries
(codes of Haardt, 1994 and Poutanen and Svensson, 1996) for the same
value of $\theta\simeq$~0.7 and for a cut--off power law spectrum ({\sc
pexrav} model of {\sc xspec}) with a e--folding energy $E_{\rm
c}=2kT_{\rm e}$=720~keV, as a first order approximation to Comptonization
spectral models (for $\tau \lta 1$). For the sphere, the soft photons are
supposed to be emitted isotropically at the center of the sphere, whereas
they come from the bottom for the slab and the hemisphere
configurations. In each case, the optical depths have been chosen so as
to produce approximatively the same spectral index in the 2-10 keV X--ray
range ($\tau$=0.09, 0.16 and 0.33 for the slab, hemisphere and sphere
geometry, respectively). We see that the spectra are quite different at
medium - high energy ($E \gta 10$ keV). In the slab and hemisphere cases,
they can be approximately described by broken power laws, the energy of
the break $E_{\rm break}$ rougly lying between the second and the third
scattering order peaks (Haardt, 1993). The slope at low energies is flat
due to the deficiency of photons caused by the anisotropy of the first
scattering and is higly angle-dependent. On the contrary, since the
higher scattering orders are almost isotropic, the slope above the break
is almost angle-independent. Due to the presence of this break,
Comptonization spectra (in anisotropic geometry) are softer for $E \gta
E_{\rm break}$ than the cut-off power law one.

\subsection{Comptonization model versus cut-off power law}
We show in Fig. 1b the best fit models derived using Comptonization (in
slab geometry) and a cut--off power law . The two models require a
different normalization for the reflection component (larger for the
slab) and are roughly in agreement below $ 200$ keV, the upper energy end
of our data. However they differ by up to a factor 10 near 500 keV, since
the cut--off energy required by the (harder) power law model is lower
than that required by the (intrinsically curved) slab Comptonization
spectrum (see Table 1).

\begin{figure}
\hspace*{-2cm}
\begin{tabular}{ll}
\begin{minipage}{7cm}
\psfig{file=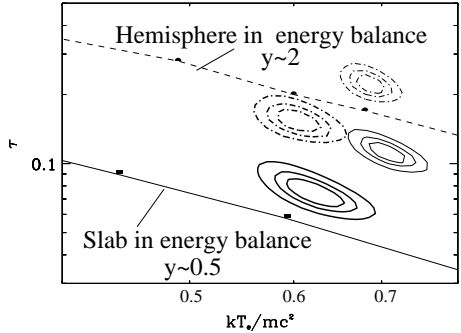, width=7cm,angle=0}
\end{minipage} &
\begin{minipage}{7cm}
\caption{Solid and dot-dashed contour plots of $\tau$ vs. $\Theta$ in the
low and high states for the slab and hemisphere configurations
respectively. The thick contours correspond to the high states. We have
also over-plotted the $\Theta-\tau$ relations predicted when energy
balance is achieved with the corresponding (rougly constant) value of the
Compton parameter (from Stern et al. 1995).}
\end{minipage}
\end{tabular}
\end{figure}

\begin{table}
\begin{tabular}{ccccccc}
\hline
State&Geom.&$\Theta$&$\tau$&R&$\gamma$&$\chi^{2}/dof$  \\
\hline
Low&Slab & 0.70$_{-0.02}^{+0.04}$ & 0.12$_{-0.02}^{+0.01}$ & 1.0$\pm 
0.2$ & -& 90/114\\
 &Hemi. & 0.68$_{-0.02}^{+0.02}$ & 0.22$_{-0.02}^{+0.03}$ & 1.9$\pm 
0.3$ & -& 80/114\\
 &{\sc pexrav} & 0.11$_{-0.02}^{+0.05}$ & - & 0.5$\pm 
0.2$ & 1.55$_{-0.02}^{+0.02}$& 93/114\\
\hline
High &Slab & 0.62$_{-0.04}^{+0.04}$ & 0.07$_{-0.01}^{+0.02}$ & 1.6$\pm 
0.5$ & -& 137/145\\
 &Hemi. & 0.61$_{-0.03}^{+0.02}$ & 0.15$_{-0.02}^{+0.03}$ & 2.7$\pm 
0.6$ & -& 143/145\\
 &{\sc pexrav} & 0.16$_{-0.07}^{+0.38}$ & - & 0.6$\pm 
0.4$ & 1.71$_{-0.04}^{+0.03}$& 142/145\\
\hline
\end{tabular}
\caption{Best fit parameters for Comptonization models in slab and
hemisphere geometries and for {\sc pexrav}. In this last case, we used
the approximation $\theta=E_c/2m_ec^2$ (valid for $\tau \lta 1$)}
\vspace*{-0.5cm}
\end{table}
\subsection{Geometry}
\label{geometry}
Both slab and hemisphere geometries give acceptable fits to the average
data (cf. Table 1). The derived parameters are not far from theoretical
expectations based on simple energy balance arguments (cf. Fig. 2). For
the slab geometry, the data suggest that the hot gas is {\it photon
starved}, i.e., it is undercooled. For the hemispherical geometry the
parameters are consistent with the energy balance condition but the
required normalization of the reflection component is too large
(cf. Table 1). This may suggest that the real geometry is intermediate
and/or that the physical situation is more complex possibly involving a
non uniform corona and deviations from strict energy balance (see Malzac
this meeting).

\subsection{Variability}
Independently of geometry, the low--to--high state transition (the
low/high state being the state outside/during the flare) clearly
indicates a change of the Compton parameter, i.e., of the
Comptonized--to--soft luminosity ratio (cf. Fig. 2). It seems to be most
probably due to an increase of the cooling rate, rather than to a
decrease of the heating rate, since we observe a pivoting at high
energies of the continuum in the two states (cf. P99). If this
interpretation is correct, then the spectral softening in the high state
is very naturally explained by a drop of the corona temperature,
ultimately due to an increase of the UV--EUV soft photon flux.

\section{Conclusions}
This \BS observation of NGC 5548 allowed us to show that i) the
temperature $kT_{\rm e}$ of the Comptonizing plasma can be largely
underestimated (up to a factor of 7 here) when derived from simple power
law models with high energy cut-off; ii) the data are well fitted by a
plane parallel corona model with an inclination angle of $30^{\circ}$, a
soft photon temperature of 5 eV, a hot plasma $kT_{\rm e}\simeq$ 360~keV
and an optical depth $\tau\simeq 0.1$. The latter values suggest however
that the hot Comptonizing gas, if in the shape of slab, is not in energy
balance. A better agreement is obtained with an hemispherical geometry;
iii) the change of state during the central part of the run clearly
indicates a variation of the Compton parameter $y$, which could be due,
as suggested by the data, to an increase of the cooling.

\begin{acknowledgements}
We gratefully acknowledge J. Poutanen for providing us his code. This
work was supported in part by the EC under contract number
ERBFMRX-CT98-0195 ( TMR network "Accretion onto black holes, compact
stars and protostars")
\end{acknowledgements}

\end{document}